\documentclass[12pt]{iopart}

\usepackage{iopams}  

\usepackage{hyperref}
\usepackage{graphicx}

\usepackage{tikz}
\usetikzlibrary{calc}
\usetikzlibrary{chains}
\usetikzlibrary{scopes}
\usetikzlibrary{shapes.geometric}
\usetikzlibrary{trees}
\usetikzlibrary{topaths}
\usetikzlibrary{positioning}

\renewcommand{\|}{\big|}
\renewcommand{\>}{\big\rangle}

\def\be{\begin{equation}}
\def\ee{\end{equation}}
\def\ba{\begin{eqnarray}}
\def\ea{\end{eqnarray}}

\begin{document}

%\title[Enumerating $q$-colourings of random planar graphs]{Enumerating $q$-colourings of random planar graphs using a tree-decomposed transfer matrix}

\title[Tree-decomposed transfer matrix]{A tree-decomposed transfer matrix for computing exact Potts model partition functions for arbitrary graphs, with applications to planar graph colourings}

\author{Andrea Bedini$^{1,2}$\footnote{Present address: MASCOS,
    Department of Mathematics and Statistics, The University of
    Melbourne, VIC, 3010, Australia} and Jesper Lykke Jacobsen$^{3}$}

\address{${}^1$ Dipartimento di Fisica, Universit\`a degli
  Studi di Milano, I-20133, Milano, Italy}
\address{${}^2$ INFN, Sezione di Milano, I-20133, Milano, Italy}
\address{${}^3$ LPTENS, \'Ecole Normale Sup\'erieure, 24 rue Lhomond, 75231 Paris, France}

\eads{\mailto{andrea.bedini@mi.infn.it}, \mailto{jesper.jacobsen@ens.fr}}

\begin{abstract}

  Combining tree decomposition and transfer matrix techniques provides
  a very general algorithm for computing exact partition functions of
  statistical models defined on arbitrary graphs.  The algorithm is
  particularly efficient in the case of planar graphs.  We illustrate
  it by computing the Potts model partition functions and chromatic
  polynomials (the number of proper vertex colourings using $Q$
  colours) for large samples of random planar graphs with up to
  $N=100$ vertices. In the latter case, our algorithm yields a
  sub-exponential average running time of $\sim \exp(1.516 \sqrt{N})$, a substantial
  improvement over the exponential running time $\sim \exp(0.245 N)$ provided by the hitherto
  best known algorithm.  We study the statistics of chromatic roots of
  random planar graphs in some detail, comparing the findings with
  results for finite pieces of a regular lattice.

\end{abstract}

\vspace{2pc}
\noindent{\it Keywords}: Tree decomposition, transfer matrix,
chromatic polynomial, random planar graphs
% \maketitle

\section{Introduction}

A typical problem in statistical physics is to compute the partition
function $Z$ of some model with short-ranged interactions between
discrete degrees of freedom defined on the vertices of some graph.
The partition function is a weighted sum over the states of these
degrees of freedom.  If the graph is a regular lattice (i.e., consists
of a large number $M$ of identical layers) one usually rewrites $Z$ in
terms of matrix elements of $T^M$, where $T$ is a transfer matrix (or
imaginary-time evolution operator in an equivalent path integral
formulation) that corresponds to the addition of a single layer to the
lattice. This suggests solving the problem by diagonalising $T$, so
that, in some sense, one has substituted algebraic complexity for
combinatorial complexity.  When the lattice is planar, the exact
diagonalisation of $T$ in the limit of an infinitely large system (or
thermodynamic limit) can in many cases be achieved by using the
powerful tools of integrability \cite{Baxterbook}. Alternatively, there exists many
efficient means of diagonalising $T$ numerically for rather large
systems.

In many applications one is however interested in models defined on
graphs that are not regular, but incorporate some element of
randomness. One must then distinguish whether this randomness is of
the annealed or the quenched type. For the annealed disorder, $Z$ is a
double sum over the graphs and the vertex degrees of freedom. In the
case of planar graphs, this double sum can in many cases be evaluated
analytically by random matrix techniques \cite{Mehta}. But arguably, the physically
most relevant scenario is that of quenched disorder, where one would
typically need to average the free energy $\log Z$, not just $Z$, over
the disorder realisations.

Clearly, a maximal way to deal with quenched randomness would be to
evaluate $Z$ independently for each graph in the sample. This is
generally not possible analytically, even for planar graphs. The
purpose of this paper is to exhibit an efficient algorithm that
evaluates $Z$ exactly for an arbitrarily given graph. This algorithm
applies to a rather large class of models of statistical physics, as
outlined above, and it does not require the graph to be planar. Rather
than delving into this generality---we shall briefly come back to the
issue in the discussion---we have here chosen to focus on one
significant specific application, namely the evaluation of the Potts
model \cite{Potts52} partition function $Z_G(Q,v)$ (which, after a
change of variables, equals the Tutte polynomial \cite{Tutte54} known
in graph theory) for a given planar graph $G$. We are particularly
interested in the special case $v=-1$, where $Z_G(Q,v) \equiv \chi_G(Q)$ equals the
number of $Q$-colourings (chromatic polynomial) of $G$. We now outline
the physical motivation for studying this problem, before ending the
introduction by giving a brief account on the working principle of our
algorithm and its performance (computational complexity).

\bigskip

The $Q$-colouring problem consists in assigning to each of the
vertices of a graph $G$ any one of $Q$ different colours, in such a
way that adjacent vertices carry different colours. Each such
assignment is known as a proper vertex colouring.  The $Q$-colouring
problem arises within physics in studies of frustrated
antiferromagnetic spin models, and in spin glass theory in particular
\cite{Mulet02}.  The number of possible colourings (possibly zero) can
be shown \cite{Birkhoff12,Whitney32} to be a polynomial in $Q$, known
as the chromatic polynomial $\chi_G(Q)$ of the graph. In particular it
makes sense---and is useful---to generalise the original counting
problem and to study the properties of $\chi_G(Q)$, considered as a
polynomial of a formal variable $Q$.

The history of the $Q$-colouring problem is long and interesting, and
we refer the reader to \cite{paper1} for an extensive list of
references. The case where $G$ is a {\em planar} graph has attracted
particular interest. A well-known result is then the four-colouring
theorem \cite{AppelHaken76} which states that $\chi_G(4) > 0$ for any
planar $G$. Other results exploit the extension of $\chi_G(Q)$ to
non-integer values of $Q$. One interesting question is whether there
exists, in the planar case, some $Q_{\rm c}$ so that $\chi_G(Q) > 0$
for all $Q \in [Q_{\rm c},\infty)$. This statement has been
established as a theorem \cite{BL46} for $Q_{\rm c} = 5$, and the
so-called Birkhoff-Lewis conjecture \cite{BL46} that this extends to
$Q_{\rm c} = 4$ is widely believed to be true.

The case where $G$ is a {\em regular} planar lattice has also been
intensively studied. In particular, the location and properties of
chromatic roots, $\chi_G(Q) = 0$, in the complex $Q$-plane has been
studied for a variety of lattices and boundary conditions (see
\cite{Baxter87,paper1,paper2,paper3,paper4,torus} and references
therein). Some of the mechanisms responsible for the generation of
real chromatic roots in the region $Q \in [0,4]$, close to or exactly
at the so-called Beraha numbers $B_k = \big(2 \cos(\pi/k) \big)^2$
with $k \ge 2$ integer, have even been understood analytically
\cite{Saleur90,Saleur91}. One should also mention that there exists a
family of planar graphs---planar triangulations actually---with real
chromatic roots converging to $Q=4$ from below \cite{Royle05}, meaning
that the value of $Q_{\rm c}$ in the Birkhoff-Lewis conjecture cannot
be lowered further.

One issue which has been left largely unanswered by these studies is
the distribution of the location of (real and complex) chromatic roots
for ensembles of randomly chosen planar graphs. This situation appears
to be the most interesting from the point of view of the physics of
disordered frustrated systems. From the mathematical side, it has been
proven \cite{Sokal04} that chromatic roots are {\em dense} in the
complex plane (except maybe in the disc $|Q-1| < 1$) for a special
class of planar graphs (generalised theta graphs). However, this might
not have much to do with the physical question of the chromatic roots
of a {\em typical} planar graph. One would also want to know whether
typical roots accumulate at the Beraha numbers, and which graph
characteristics (connectivity, average coordination number,\ldots)
might be responsible for the location of roots.

\bigskip

In order to elucidate the amazingly intricate behaviour of chromatic
roots in the limit of large graphs, it is clearly desirable to have
efficient means of computing $\chi_G(Q)$. Let us recall that in
theoretical computer science, an important outstanding question is
whether the class P of decision problems that can be {\em answered} in
polynomial time coincides with the class NP of problems for which a
proposed answer can be {\em verified} in polynomial time. NP-complete
problems are those to which any problem in NP can be reduced in
polynomial time. At present, no polynomial-time algorithm has been
found for any of the thousands of known NP-complete problems and it is
hence widely believed that P $\neq$ NP. Likewise, one can define a
counting analogue of NP, denoted by \#P, as the class of enumeration
problems in which the objects being counted are the possible answers
accepted by an NP machine. If an NP problem is often of the form ``Are
there any solutions that satisfy certain constraints?'', the
corresponding \#P problems ask ``how many'' rather than ``are there
any''. The class \#P-complete of the hardest problems in \#P is
defined likewise. Clearly, \#P-complete problems are at least as hard
as NP-complete problems. It is known \cite{Welsh} that the counting of
proper vertex colourings is a \#P-complete problem for $Q=3$, and the
same is then true {\em a fortiori} for the computation of $\chi_G(Q)$,
let alone $Z_G(Q,v)$ \cite{Jaeger90}, for generic values of $Q$ and
$v$.

In practice this means that any algorithm computing $\chi_G(Q)$ can be
expected to have a running time that increases exponentially with the
number of vertices $N$. However, lowering the coefficient of the
exponent can still make a huge difference for studying the issues
outlined above. One central goal of this paper is to make a
substantial step in that direction, by lowering the average running time $\sim
\exp(0.245 N)$ of the previously best known algorithm
\cite{HaggardRoyle} to $\sim \exp(1.516 \sqrt{N})$, as illustrated in
Fig.~\ref{fig:comparison} below. The improvement from exponential to sub-exponential asymptotics is not only important from a theoretical point of view. To get a rough idea of what this
means in practice, in ten seconds the algorithm \cite{HaggardRoyle}
will compute $\chi_G(Q)$ for a typical planar graph of $N=40$
vertices, whereas our algorithm can deal with $N=100$ in the same
time.

\bigskip

Algorithmic progress on \#P-complete problems related to graph theory and
network design has been made by several, usually widely separated,
communities.

On one hand, statistical physicists have shown that the relevant
partition functions can be constructed in analogy with the path
integral formulation of quantum mechanics. To this end, the
configuration of a partially elaborated graph are encoded as suitable
quantum states, and the constant-time surface is swept over the graph
by means of a time evolution operator known as the transfer
matrix. Although rarely stated, this approach is valid not only for
regular lattices but also for arbitrary graphs.

On the other hand, graph theorists have used that graphs can be
divided into ``weakly interacting'' subgraphs through a so-called tree
decomposition \cite{treedecomp,treedecompreview}, and solutions
obtained for the subgraphs can be recursively combined into a complete
solution.

The principle underlying the algorithmic progress to be described in
this paper is that tree decomposition and transfer matrix methods can
be combined in a very natural way. The main idea is that the tree
decomposition is compatible with a recursive generalisation of the
time evolution concept. Borrowing ideas from quantum field theory, the
combination of partial solutions is obtained by the fusion of suitable
state spaces. The resulting algorithm works on any graph, and can
readily be adapted to many other problems of statistical physics, by
suitable modifications of the state spaces and the fusion procedure.

In particular, we will apply this technique to the problem of
computing the chromatic polynomial on planar graphs, obtaining exact
solutions, for graphs with $N \simeq 100$ vertices, in only a few
seconds.

\bigskip

The layout of the paper is as follows. In section \ref{sec:potts} we
briefly recall the relation between the Potts model partition function
$Z_G(Q,v)$ and the chromatic polynomial $\chi_G(Q)$. In section
\ref{sec:algo} we present our algorithm and discuss its performance.
The results for chromatic roots of planar graphs are given and discussed
in section \ref{sec:res}. Finally, we give our conclusions in section
\ref{sec:concl}.

\section{Potts model and vertex colourings}
\label{sec:potts}

We first recall the relation between the vertex colouring problem and
the Potts model.
Consider a graph $G=(V,E)$ with vertices $V$ and edges $E$, and let
$\sigma_i = 1,2,\ldots,Q$ be the colour of vertex $i \in V$. Then
\be
 Z_G
% = \sum_{\sigma} {\rm e}^{K \sum_{(ij) \in E} \delta(\sigma_i,\sigma_j)}
= \sum_{\sigma} \prod_{(ij) \in E} {\rm e}^{K \delta(\sigma_i,\sigma_j)}
 \label{Potts}
\ee
is the partition function of the Potts model on $G$. The Kronecker
delta $\delta(\sigma_i,\sigma_j)=1$ if $\sigma_i=\sigma_j$, and 0
otherwise. Inserting the obvious identity ${\rm e}^{K
  \delta(\sigma_i,\sigma_j)} = 1 + v \delta(\sigma_i,\sigma_j)$, with
$v = {\rm e}^K-1$, and expanding out the product we obtain the
Fortuin-Kasteleyn expansion \cite{FK}
\be
 Z_G(Q,v) = \sum_{A \subseteq E} v^{|A|} Q^{k(A)} \,,
 \label{FK}
\ee
where $k(A)$ is the number of connected components in the subgraph
$G'=(V,A)$. Obviously, in the antiferromagnetic limit $K \to
- \infty$ (or equivalently $v \to -1$) the only surviving
configurations are proper $Q$-colourings of the graph $G$. Indeed, the
special case $\chi_G(Q) = Z_G(Q,-1)$ is a polynomial in $Q$, known as
the {\em chromatic polynomial}, and equals the number of vertex
colourings.

\section{Algorithm}
\label{sec:algo}

\subsection{Transfer matrix}

The computation of the partition function $Z_G(Q,v)$ using the expansion
(\ref{FK}) has been described in \cite{BN82,paper1} for the case where
$G$ is a finite piece of a {\em regular} lattice (notably, but not
exclusively, a planar one).

We first describe how this traditional transfer matrix method extends
to the case where $G$ is an arbitrary graph---i.e., not necessarily
part of a regular lattice, nor necessarily planar. In short, the
combined action of linear operators builds a superposition of all
configurations appearing in the partition funciton with their correct
statistical (Boltzmann) weight entering as coefficients.  To better
illustrate this procedure, consider the following example graph $G$
\begin{equation}
\begin{tikzpicture}[baseline=(current bounding box.center)]
	\path (0,0) node (a) {1}
		(0,-1) node (b) {2}
		(1,0) node (c) {3}
		(1,-1) node (d) {4}
		( $(c) + (-30:1)$ ) node (e) {5}
		( $(c) + (60:1)$ ) node (f) {6}
		( $(e) + (60:1)$ ) node (g) {7}
		( $(d) + (-60:1)$ ) node (h) {8}
		( $(e) + (-60:1)$ ) node (i) {9}
		;
	\draw (c) -- (a) -- (b) -- (d) -- (h) -- (i)
		-- (e) -- (g) -- (f) -- (c) -- (d) -- (e)
		-- (c);
\end{tikzpicture}
\label{graph}
\end{equation}
We first have to define the order $\{ v_t \}$ in which vertices will
be processed.  This order is the basis for the construction of a
``time slicing'' of the graph. With each \emph{time step} is
associated a \emph{bag} (a vertex subset) of active vertices. A vertex
becomes active as soon as one of its neighbours is processed and it stays
active until it is processed itself. Taking the vertices in
lexicographic order we obtain the following decomposition:
\begin{equation}
\begin{tikzpicture}[bag/.style={ellipse,draw,inner sep=2pt},baseline=(current bounding box.center)]
	{ [start chain,node distance=3mm,every join/.style={->}]
	\node [bag,on chain] {\textbf 1 2 3};
	\node [bag,on chain,join] {\textbf 2 3 4} ;
	\node [bag,on chain,join] {\textbf 3 4 5 6};
	\node [bag,on chain,join] {\textbf 4 5 6 8};
	\node [bag,on chain=going below,join] {\textbf 5 6 7 8 9};
	\node [bag,continue chain=going left,on chain,join] {\textbf 6 7 8 9};
	\node [bag,on chain,join] {\textbf 7 8 9};
	\node [bag,on chain,join] {\textbf 8 9};
	\node [bag,on chain,join] {\textbf 9};
	}
\end{tikzpicture}
\label{path_dec}
\end{equation}
where we wrote in bold face the vertex being processed at each time step.

Each bag has its own set of basis states consisting of the
\emph{partitions}%
\footnote{Recall that a partition of a finite set $S$ is a collection
  of nonempty, pairwise disjoint subsets of $S$ whose union is
  $S$. The subsets are called {\em blocks}.}
of the currently active vertices. For instance, in the first time
step, the basis states are the five partitions of the three-element
set $\{1,2,3\}$:
\tikzset{connectivity/.style={x=12pt,y=8pt,baseline=(1.south)}}
\begin{equation*}
\| \tikz[connectivity] {
  \foreach \x in {1,2,3}
  \draw (\x-1,0) circle (2pt) node (\x) {}
  node[below] {$\scriptstyle \x$};
}\> \,, \quad
\| \tikz[connectivity] {
  \foreach \x in {1,2,3}
  \draw (\x-1,0) circle (2pt) node (\x) {}
  node[below] {$\scriptstyle \x$};
  \draw (1.north) .. controls +(0,0.5) and +(0,0.5) .. (2.north);
}\> \,, \quad
\| \tikz[connectivity] {
  \foreach \x in {1,2,3}
  \draw (\x-1,0) circle (2pt) node (\x) {}
  node[below] {$\scriptstyle \x$};
  \draw (2.north) .. controls +(0,0.5) and +(0,0.5) .. (3.north);
}\> \,, \quad
\| \tikz[connectivity] {
  \foreach \x in {1,2,3}
  \draw (\x-1,0) circle (2pt) node (\x) {}
  node[below] {$\scriptstyle \x$};
  \draw (1.north) .. controls +(0,0.5) and +(0,0.5) .. (3.north);
}\> \,, \quad
\| \tikz[connectivity] {
  \foreach \x in {1,2,3}
  \draw (\x-1,0) circle (2pt) node (\x) {}
  node[below] {$\scriptstyle \x$};
  \draw (1.north) .. controls +(0,0.5) and +(0,0.5) .. (2.north);
  \draw (2.north) .. controls +(0,0.5) and +(0,0.5) .. (3.north);
}\> \,.
\end{equation*}
These partitions describe how the active vertices are interconnected
through $A \cap E_t$, where $E_t \subseteq E$ is the subset of edges
having been processed at time $t$. A \emph{state} is a linear
superposition of basis states.

Processing a vertex consists in processing edges connecting it to
unprocessed vertices and then deleting it. Since each edge $e \in E$
may or may not be present in $A$ we process an edge $(i,j)$ by acting
on the state with an operator of the form $\mathsf 1 + v \mathsf
J_{ij}$ where $\mathsf 1$ is the identity operator and $\mathsf
J_{ij}$ a {\em join operator}. A join operator acts on a basis state
by amalgamating the blocks containing vertices $i$ and $j$.
\begin{equation}
  \mathsf J_{ij} \, \|\tikz[connectivity] {
    \draw
    (0,0)  circle (2pt) node (i) {} node[below] {$\scriptstyle i$}
    (1,0)  circle (2pt) node (j) {} node[below] {$\scriptstyle j$};
  }\> = \| \tikz[connectivity] {
    \draw
    (0,0)  circle (2pt) node (i) {} node[below] {$\scriptstyle i$}
    (1,0)  circle (2pt) node (j) {} node[below] {$\scriptstyle j$};
    \draw (i.north) .. controls +(0,0.5) and +(0,0.5) .. (j.north);
  }\> \,, \qquad
  \mathsf J_{ij}^2 = \mathsf J_{ij} \,.
\end{equation}
Vertex deletion is defined in terms of a {\em deletion operator}
$\mathsf D_i$ that removes $i$ from the partiton and applies a factor
$Q$ (resp. 1) if $i$ was (resp. was not) a singleton.
\begin{eqnarray}
  \mathsf D_i \, \| \tikz[connectivity] {
    \draw
    (0,0) circle (2pt) node (i) {} node[below] {$\scriptstyle i$}
    (1,0) circle (2pt) node (j) {} node[below] {$\scriptstyle j$};
  } \cdots \> & = Q \, \| \tikz[connectivity] {
    \draw
    (0,0) circle (2pt) node (j) {} node[below] {$\scriptstyle j$};
  } \cdots \> \,, \\
  \mathsf D_i \, \| \tikz[connectivity] {
    \draw
    (0,0) circle (2pt) node (i) {} node[below] {$\scriptstyle i$}
    (1,0) circle (2pt) node (j) {} node[below] {$\scriptstyle j$};
    \draw (i.north) .. controls +(0,0.5) and +(0,0.5) .. (j.north);
  } \cdots \> & = \| \tikz[connectivity] {
    \draw
    (0,0) circle (2pt) node (j) {} node[below] {$\scriptstyle j$};
  } \cdots \> \,.
\end{eqnarray}
For example, in (\ref{path_dec}), processing the first bag means
computing the following composition
\begin{equation*}
%  \label{eq:1}
  \mathsf D_1 \,
  (\mathsf 1 + v \mathsf J_{12}) \,
  (\mathsf 1 + v \mathsf J_{13}) \,
  \| \tikz[connectivity] \foreach \x in {1,2,3}
  \draw (\x-1,0) circle (2pt) node (\x) {}
  node[below] {$\scriptstyle \x$};
  \> \,,
\end{equation*}
which gives
\begin{equation*}
  (Q + 2v) \, \|
  \tikz[connectivity] {
    \draw (0,0) circle (2pt) node (2) {}
          node[below] {$\scriptstyle 2$}
          (1,0) circle (2pt) node (3) {}
          node[below] {$\scriptstyle 3$}; 
  }\> + v^2 \|
  \tikz[connectivity] {
    \draw (0,0) circle (2pt) node (2) {}
                node[below] {$\scriptstyle 2$}
          (1,0) circle (2pt) node (3) {}
                node[below] {$\scriptstyle 3$};
    \draw (2.north) .. controls +(0,0.5) and +(0,0.5) .. (3.north);
  }\> \,,
\end{equation*}
concluding the first time step.

When a new active vertex is encountered it is inserted, as a
singleton, in each partition composing the current state. After
processing the last bag, the complete partition function (\ref{FK}) is
obtained as the coefficient of the empty partition resulting from the
deletion of the last active vertex.

At each step, the time and memory requirements are determined by the
{\em bag size} $n$ which is the number of vertices simultaneously
active. If the graph is planar, the number of partitions to be
considered is at most the Catalan number $C_n$, i.e., the number of
non-crossing partitions. The corresponding generating function is
\begin{equation}
 C(z) = \sum_{n=0}^\infty C_n z^n = \frac{1-\sqrt{1-4z}}{2z} \,.
 \label{Catalan}
\end{equation}
If the graph is not planar, the number of partitions is at most the Bell
number $B_n$, with generating function
\begin{equation}
 B(z) = \sum_{n=0}^\infty B_n \frac{z^n}{n!}  = \exp({\rm e}^z-1) \,.
\end{equation}
We have $C_n = 4^n n^{-3/2} \pi^{-1/2} [1+O(1/n)]$, whereas $B_n$
grows super-exponentially.%
\footnote{Note that the restriction to non-crossing partitions holds
  true even if the ordering of the vertices, such as in
  (\ref{path_dec}), does not respect the planar embedding (\ref{graph})
  of the whole graph (which will in general be unknown). Because we
  choose to store partitions independently of their crossing property,
  it has not been relevant to be able to explicitly keep track of the
  planar embedding. All that matters is that the number of partitions
  does not exceed $C_n$, and this is true since such a non-crossing
  representation could be obtained ``by hand'' by inspecting the
  planar embedding at each stage of the transfer process.}

\subsection{Tree decomposition}

It turns out that the decomposition (\ref{path_dec}) of $G$ is a
special case of a more general construction. By definition, a {\em
  tree decomposition} \cite{treedecomp,treedecompreview} of a graph $G
= (V,E)$ is a collection of bags, organised as a tree (a connected
graph with no cycles), and satisfying the following requirements:
\begin{enumerate}
 \item For each $i \in V$, there exists a bag containing $i$; 
 \item For each $(ij) \in E$, there exists a bag containing both $i$ and $j$;
 \item For any $i \in V$, the set of bags containing $i$ is connected in the
       tree.
\end{enumerate}
The previous decomposition (\ref{path_dec}) is just a special case of
a tree decomposition (a {\em path decomposition}). As an example of
the general construction, applied to (\ref{graph}), consider
\begin{equation}
\begin{tikzpicture}[bag/.style={ellipse,draw,inner sep=2pt},baseline=(current bounding box.center)]
  { [start chain,node distance=3mm,every join/.style={->}]
    \node [bag,on chain] {1 2 3};
    \node [bag,on chain,join] {2 3 4};
    \node [bag,on chain,join] {3 4 5};
    
    { [start branch=2 going above right,every join/.style={<-}]
      \node [bag,on chain,join] {3 5 6};
      \node [bag,continue chain=going right,on chain,join] {5 6 7};
    }		
    { [start branch=3 going below right,every join/.style={<-}]
      \node [bag,on chain,join] {4 5 8};
      \node [bag,continue chain=going right,on chain,join] {5 8 9};
    }
  }
\end{tikzpicture}
\label{tree_decomp}
\end{equation}
where the arrows form the unique path that connects each bag to the
central one (the \emph{root} of the tree).

The advantage of working with tree instead of path decompositions
relies on the fact that in the former case a decomposition with
smaller bags can be obtained (the latter being just a special
case). Therefore, the number of states one has to keep track of is
exponentially smaller, and the gain is significant.

The transfer matrix approach can be adapted naturally to this new
general setting: Properties (i)--(ii) guarantee that each edge and
vertex are processed within a definite bag. Property (iii) implies that
each vertex has a definite life time in the recursion, its
insertion and deletion being separated by the processing of all edges
incident on it.

In this new version the algorithm starts from the root of the tree,
which can be chosen arbitrarily, and runs through the tree
recursively. Alternatively, this can be thought of as starting from
the leaves and building up the tree inductively.

Consider first the case of a parent bag $P$ with only one daughter bag
$D$. Going up from $D$ to $P$ in the recursive step involves deleting
vertices $D \setminus P$, inserting vertices $P \setminus D$ and
finally processing edges in $P$.  A tree decomposition does not
specify when an edge $e=(ij)$ must be processed.  A simple recipe
would be to process $e$ as soon as one encounters a bag containing
both $i$ and $j$. However we note that this freedom of choice can be
exploited to optimise the algorithm (see below).

\subsection{Fusion}

We now discuss the case when a parent bag $P$ has several daughters
$D_\ell$ with $\ell=1,2,\ldots,d$. In this case, vertex deletions and
insertions are followed by a special {\em fusion} procedure. Suppose
first $d=2$, and let ${\cal P}_1$ be a partition of $D_1 \cap P$ with
weight $w_1$, and ${\cal P}_2$ a partition of $D_2 \cap P$ with weight
$w_2$. The fused state is then
\begin{equation*}
 {\cal P}_1 \otimes {\cal P}_2 = {\cal P}_1 \vee {\cal P}_2 \,,
\end{equation*}
and it occurs with weight $w_1 w_2$.  Here $\vee$ denotes the join
operation in the partition lattice.%
\footnote{We recall that a partial order of the partitions follows
  from defining a partition ${\cal P}_a$ to be a {\em refinement} of
  ${\cal P}_b$, and we write ${\cal P}_a \preceq {\cal P}_b$, provided
  that each block in ${\cal P}_a$ is a subset of some block in ${\cal
    P}_b$.  With this partial order, the set of partitions form a {\em
    lattice}, in the sense that any two partitions ${\cal P}_1$ and
  ${\cal P}_2$ possess a largest lower bound called {\em meet} and
  denoted ${\cal P}_1 \wedge {\cal P}_2$ and a smallest upper bound
  called {\em join} and denoted ${\cal P}_1 \vee {\cal P}_2$.  The
  meet ${\cal P}_1 \wedge {\cal P}_2$ has as blocks all nonempty
  intersections of a block from ${\cal P}_1$ with a block from ${\cal
    P}_2$. The blocks of the join ${\cal P}_1 \vee {\cal P}_2$ are
  the smallest subsets which are exactly a union of blocks from both
  ${\cal P}_1$ and ${\cal P}_2$.  In the partition lattice, the bottom
  (or finest) element is the unique partition in which each block has
  size 1 (the all-singleton partition), and the top (or coarsest)
  element is the unique partition in which there is a single block
  (the all-connected partition).}  For definiteness, let us explain
how this can be computed in practice.  First choose some set $E_1$ so
that
\begin{equation}
 {\cal P}_1 = \left( \prod_{e \in E_1} {\mathsf J}_e \right) {\cal S}_1 \,,
 \label{eq:decomposition}
\end{equation}
where ${\cal S}_1$ is the all-singleton partition of $D_1 \cap P$.
Note that it is not necessary to choose $E_1$ as a subset of $E$,
although this can always be done.  Since $\mathsf J_e \mathsf J_{e'} =
\mathsf J_{e'} \mathsf J_e$ the order in the above product is
irrelevant. Similarly choose $E_2$ for ${\cal P}_2$. The fused state
can then be constructed explicitly as
\begin{equation*}
 {\cal P}_1 \otimes {\cal P}_2 = \left( \prod_{e \in E_1 \cup E_2}
 {\mathsf J}_e \right) {\cal S}_{12} \,,
\end{equation*}
where ${\cal S}_{12}$ is the
all-singleton partition of $(D_1 \cup D_2) \cap P$.  For $d>2$
daughters, the complete fusion can be accomplished by fusing $D_1$
with $D_2$, then fusing the result with $D_3$, and so on.

Let us illustrate the fusion procedure for $G$ with the tree
decomposition (\ref{tree_decomp}). After processing the two left-most
bags and deleting vertex $2$, the propagating state is
\begin{equation}
  \omega_{1} \|
  \tikz[connectivity] { \draw
    (0,0) circle (2pt) node (3) {} node[below] {$\scriptstyle 3$}
    (1,0) circle (2pt) node (4) {} node[below] {$\scriptstyle 4$}; 
  }\>
  + \omega_{2} \|
  \tikz[connectivity] { \draw
    (0,0) circle (2pt) node (3) {} node[below] {$\scriptstyle 3$}
    (1,0) circle (2pt) node (4) {} node[below] {$\scriptstyle 4$}
    (3.north) .. controls +(0,0.5) and +(0,0.5) .. (4.north);
  }\> =
  (\omega_{1} + \omega_{2} {\mathsf J}_{34})
  \| \tikz[connectivity] { \draw
    (0,0) circle (2pt) node (3) {} node[below] {$\scriptstyle 3$}
    (1,0) circle (2pt) node (4) {} node[below] {$\scriptstyle 4$}; 
  }\>,
  \label{eq:3}
\end{equation}
where $\omega_{1} = Q^2 + 3 v (Q+v)$ and $\omega_{2}
= Q^{2} v + 3 Q v^{2} + 4 v^{3} + v^{4}$.  By symmetry, the same
result is obtained for the two top right bags (replacing $4$ by $5$)
and for the two bottom right bags (replacing $3$ by $5$).%
\footnote{Needless to say, the fusion procedure will work also in cases
where the graph possesses no such symmetries. In our example, we have
chosen $G$ as a rather symmetric graph in order to keep the actual
computations as simple as possible.}
The fused
state arriving in the central bag is then
\begin{eqnarray}
 (\omega_1+\omega_2 {\mathsf J}_{34}) &
 (\omega_1+\omega_2 {\mathsf J}_{35})
 (\omega_1+\omega_2 {\mathsf J}_{45})
 \| \tikz[connectivity] { \draw
   (0,0) circle (2pt) node (3) {} node[below] {$\scriptstyle 3$}
   (1,0) circle (2pt) node (4) {} node[below] {$\scriptstyle 4$}
   (2,0) circle (2pt) node (5) {} node[below] {$\scriptstyle 5$}; 
 }\>  \nonumber \\
 &= \omega_1^3
 \| \tikz[connectivity] { \draw
   (0,0) circle (2pt) node (3) {} node[below] {$\scriptstyle 3$}
   (1,0) circle (2pt) node (4) {} node[below] {$\scriptstyle 4$}
   (2,0) circle (2pt) node (5) {} node[below] {$\scriptstyle 5$}; 
 }\> 
 + (3 \omega_1 \omega_2^2 + \omega_2^3) 
 \| \tikz[connectivity] { \draw
   (0,0) circle (2pt) node (3) {} node[below] {$\scriptstyle 3$}
   (1,0) circle (2pt) node (4) {} node[below] {$\scriptstyle 4$}
   (2,0) circle (2pt) node (5) {} node[below] {$\scriptstyle 5$}
   (3.north) .. controls +(0,0.5) and +(0,0.5) ..
   (4.north) .. controls +(0,0.5) and +(0,0.5) .. (5.north);
 }\>
 \nonumber \\
 & + \omega_1^2 \omega_2 \big(
 \| \tikz[connectivity] { \draw
   (0,0) circle (2pt) node (3) {} node[below] {$\scriptstyle 3$}
   (1,0) circle (2pt) node (4) {} node[below] {$\scriptstyle 4$}
   (2,0) circle (2pt) node (5) {} node[below] {$\scriptstyle 5$}
   (3.north) .. controls +(0,0.5) and +(0,0.5) .. (4.north);
 }\>
 + \| \tikz[connectivity] { \draw
   (0,0) circle (2pt) node (3) {} node[below] {$\scriptstyle 3$}
   (1,0) circle (2pt) node (4) {} node[below] {$\scriptstyle 4$}
   (2,0) circle (2pt) node (5) {} node[below] {$\scriptstyle 5$}
   (4.north) .. controls +(0,0.5) and +(0,0.5) .. (5.north);
 }\>
 + \| \tikz[connectivity] { \draw
   (0,0) circle (2pt) node (3) {} node[below] {$\scriptstyle 3$}
   (1,0) circle (2pt) node (4) {} node[below] {$\scriptstyle 4$}
   (2,0) circle (2pt) node (5) {} node[below] {$\scriptstyle 5$}
   (3.north) .. controls +(0,0.5) and +(0,0.5) .. (5.north);
 }\> \big)
 \, ,
  \label{eq:2}
\end{eqnarray}
from which the result $Z_G(Q,v)$ follows upon deleting vertices 3,4,5.

\subsection{Pruning}

Problem specific features can be exploited to reduce further the
number of basis states to be considered. As an example of this, note
that in the colouring case ($v=-1$) the operator $\mathsf O_{ij} = 1 +
v \mathsf J_{ij}$ associated with an edge $(ij)$ is a projector,
$\mathsf O_{ij}^2 = \mathsf O_{ij}$, and it annihilates the subspace
of partitions where $i$ and $j$ are connected. It follows that one can
discard basis states in which two vertices are connected, as soon as
one discovers that an edge between them is going to be processed later
within the same bag or within the parent bag. Especially before
fusions this simple trick reduces substantially the number of basis
states and thus leads to a big speed up.

\subsection{Performance}

For a planar graph, the state of a bag of size $n$ is spanned by $C_n$
basis states. (For a non-planar graph, replace $C_n$ by $B_n$.) The
memory needed by the algorithm is therefore proportional to $C_{n_{\rm
    max}}$, where $n_{\rm max}$ is the size of the largest bag. The
time needed to process one edge in a bag of size $n$ is proportional
to $C_n$.

However, most of the time is spent fusing states. For a parent $P$
with $d$ daughters $D_\ell$, the number of basis state pairs to be
fused is
\begin{equation}
 \sum_{\ell=1}^d C_{|{\cal D}_{\ell-1} \cap P|} C_{|D_\ell \cap P|} \,,
 \label{fusion}
\end{equation}
where we have set ${\cal D}_k = \cup_{\ell=1}^k D_\ell$. Each of these
elementary fusions (i.e., the joins ${\cal P}_1 \vee {\cal P}_2$) can
be done in time linear\footnote{We represent partitions as lists of
  numbers linking each participating vertex to the block it belongs
  to. The set $E_1$ in (\ref{eq:decomposition}) can be obtained in a
  single scan of this list where one keeps track of the last seen
  vertex for each block. In this representation, applying $J_e$ is
  constant time, apart from a possible linear-time procedure to bring
  the resulting list in a canonical form. This last procedure can be
  safely postponed after all the join operations, making the whole
  elementary fusion linear in the number of participating vertices.
  [Note that in \cite{Noble} the join operation is claimed to be {\em
    quadratic} in $n$.]}
in the number of participating vertices. Note that we can choose the
order of successive fusions so as to minimise the quantity
(\ref{fusion}).

It is therefore essential for the algorithm that one knows how to
obtain a good tree decomposition. The minimum of $n_{\rm max}-1$ over
all tree decompositions is known as the tree width $k$, but obtaining
this is another NP-hard problem. However, the simple algorithm {\tt
  GreedyFillIn} \cite{gfi} gives an upper bound $k_0$ on $k$ and a
tree decomposition of width $k_0$ in time {\em linear} in the number
of vertices $N$. For uniformly generated planar graphs we find that
for $N=40$---a value enabling comparison with algorithms that
determine $k$ exactly---that $k_0=k$ nearly always ($k_0=k + 1$ with
probability $\simeq 10^{-3}$). When $k_0$ is small enough that all the
partitions can fit into the computer's memory, the algorithm has
proven to be very fast with execution time in the order of seconds.

To summarise, suppose that we wish to compute $Z_G(Q,v)$ or $\chi_G(Q)$
for a planar graph of $N$ vertices and $M$ edges, and that we are given
a tree decomposition of $B$ bags, with $n_{\rm max}$ being the size of
the largest bag. The number of (binary) fusions to be performed is
$F = \sum_i (d_i-1)$, where $d_i$ is the number of daughters of bag $i$.
For simplicity we assume a computational model where the operations on
the weights can be done in unit time.%
\footnote{The weights are numbers for an evaluation of $Z_G(Q,v)$
or of $\chi_G(Q)$, polynomials with integer coefficients for $\chi_G(Q)$,
and polynomials in two variables with integer coefficients for $Z_G(Q,v)$.}
For the version of the algorithm without pruning, we have therefore
shown that the worst-case running time is asymptotically proportional to
\begin{equation}
 (N+M+B) C_{n_{\rm max}} n_{\rm max} +
 F C_{n_{\rm max}}^2 n_{\rm max} \,.
 \label{worst_case}
\end{equation}
Note that this is linear in $N$ and $M$ for a fixed $n_{\rm max}$
(cf.~\cite{Noble}).

\begin{figure}
\includegraphics[width=\columnwidth]{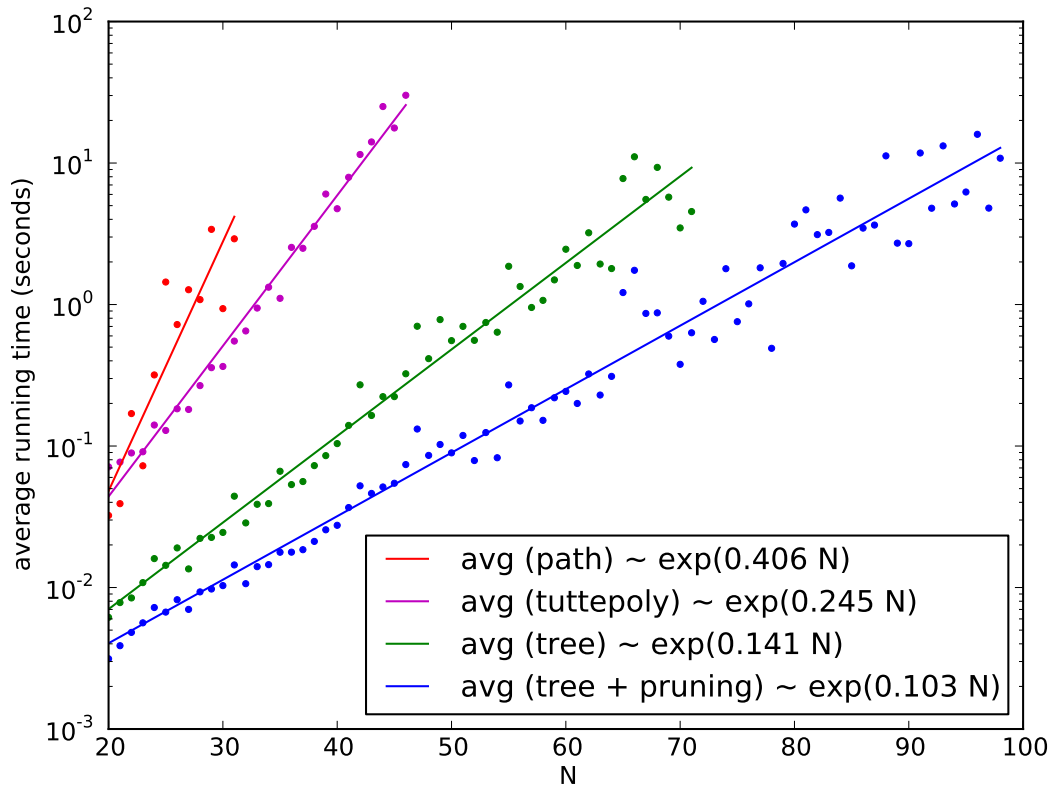}
\caption{Average running time in seconds on a random planar graph of
  size $N$ vertices. Each point is averaged over 100 graphs.}
\label{fig:comparison}
\end{figure}

We choose to test our algorithm against the one presented by Haggard
et al.\ in \cite{HaggardRoyle}. We first generated a uniform sample of 100
planar graphs for each size $N=|V|$ between 20 and 100 using Fusy's algorithm
\cite{Fusy}. We then ran four different algorithms over this sample:
the algorithm of \cite{HaggardRoyle}, our first path-based transfer matrix
algorithm, the new tree-based version algorithm and a tree-based
version using the above pruning optimisation.  Path decompositions were
obtained with a variant of the {\tt GreedyFillIn} algorithm in which
the resulting tree decomposition was forced not to branch.
Average running times are presented in Fig.~\ref{fig:comparison}, in the
form of effective exponential fits.

\begin{figure}
\includegraphics[width=\columnwidth]{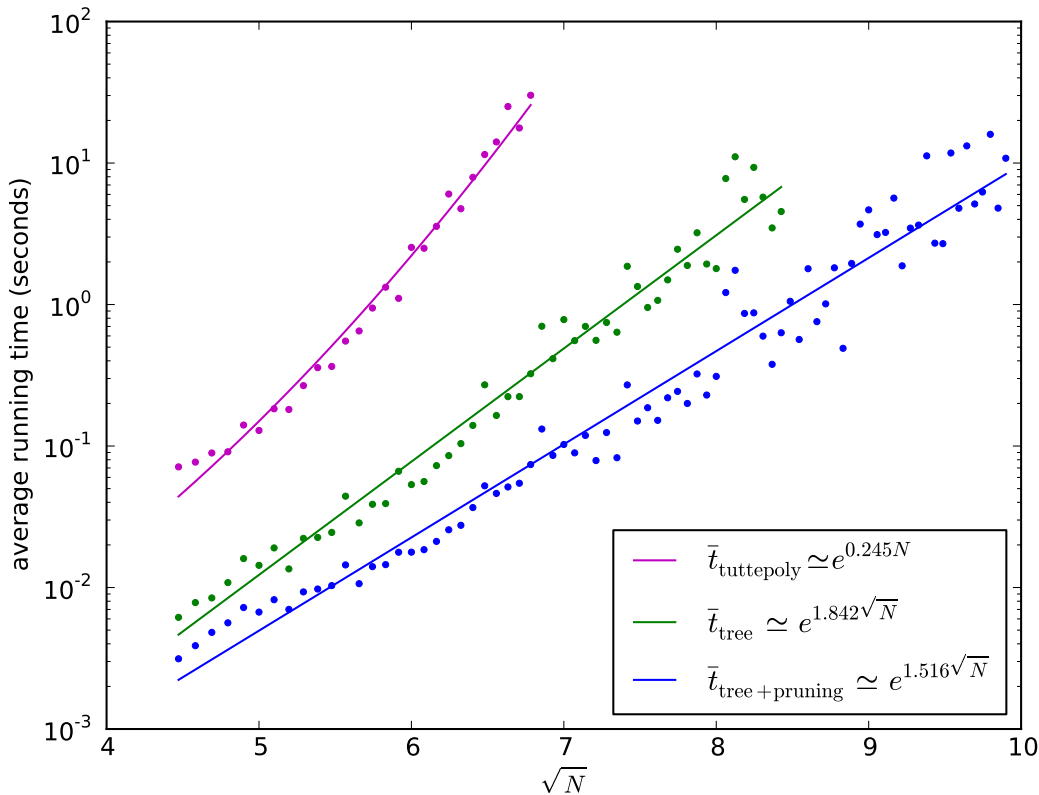}
\caption{Average running time in seconds on a random planar graph of
  size $N$ vertices, now fitted to the form $\exp(\beta \sqrt{N})$.
  The data points are identical to those of Fig.~\ref{fig:comparison}.}
\label{fig:sqrootfit}
\end{figure}

While these fits clearly confirm the efficiency of the tree decomposed
approach, they are actually misleading and hide an important
fact. Namely, the tree width of a planar graph is bounded by $\alpha
\sqrt{N}$, as follows from the famous planar separator theorem
\cite{AST}. The currently best upper bound on the constant is $\alpha
< 3.182$ \cite{Fomin}. These facts---combined with (\ref{worst_case})
and (\ref{Catalan}), and the heuristic observation of the
near-ideality of the tree decompositions obtained by the {\tt
  GreedyFillIn} heuristics---immediately imply an upper bound on the
worst-case running time on the tree-based algorithms of $16^{3.182
  \sqrt{N}} = \exp(8.822 \sqrt{N})$.  It follows that the mean running
time of these algorithms must be of the form $\exp(\beta \sqrt{N})$,
with the constants $\beta$ satisfying $\beta_{\rm tree+pruning} \le
\beta_{\rm tree}$. The fits for $\beta$, using the same data as in
Fig.~\ref{fig:comparison}, are shown in Fig.~\ref{fig:sqrootfit}.

\subsection{Technical aspects}

For completeness we give a few implementational details.  Partitions
relevant to the propagating state are kept in a hash table for fast
(amortised constant time) access.
The corresponding weights are polynomials in $Q$ whose coefficients
rapidly exceed the machine integer size $M=2^{32}$ when running on big
graphs.  To solve this issue without requiring additional memory we
used modular arithmetics: the algorithm was run many times with
coefficients computed modulo primes $p < M$, and the original
coefficients were reconstructed by the Chinese remainder theorem.

\section{Results}
\label{sec:res}

As a simple application of our algorithm we obtained the distribution
of the chromatic roots $\{ Q \,|\, \chi_G(Q)=0 \}$ for ensembles of
planar graphs of up to $N=100$ vertices. This problem is interesting
both in statistical mechanics and in graph theory
\cite{Tutte54,Saleur90,Saleur91,paper1}. As known from Lee-Yang
theory, the roots signal phase transitions.

Before turning to the actual results, it is useful to give some
results for {\em regular} planar graphs of a few hundred vertices. To
this end, we have used a particular adaptation of the traditional
(path-decomposed) transfer matrix algorithm that takes special
advantage of the regular lattice structure \cite{JLJ_series} to
compute $\chi_G(Q)$ for $L \times L$ sections of the square and
triangular lattice (the latter being considered as a square lattice
with added diagonals). The boundary conditions are free--free, in the
terminology of \cite{paper1}. The chromatic polynomials for $L=10,12$
have been checked against \cite{paper2}, while those for $L=14,16,18$
are new.

\begin{figure}
\begin{center}
\includegraphics[width=0.6\columnwidth]{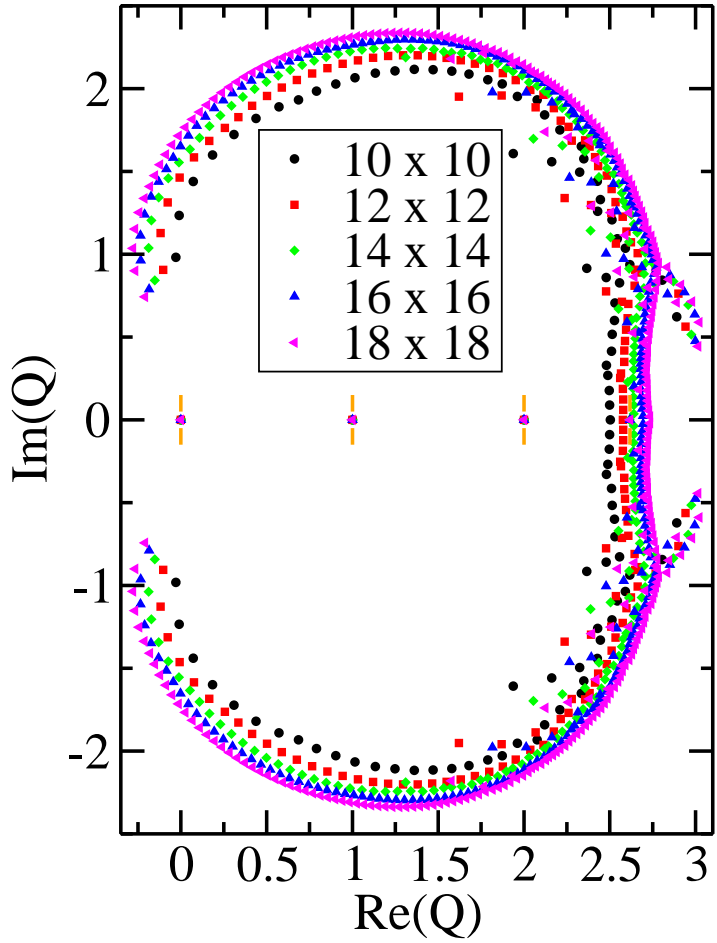}
\end{center}
\caption{Chromatic roots for $L \times L$ sections of the square
  lattice with free boundary conditions, and
  $L=10,12,14,16,18$. Beraha numbers $B_k$ with $k=2,3,4,5$ are shown
  as small vertical line segments.}
\label{fig:sq_latt}
\end{figure}

\begin{figure}
\begin{center}
\includegraphics[width=0.6\columnwidth]{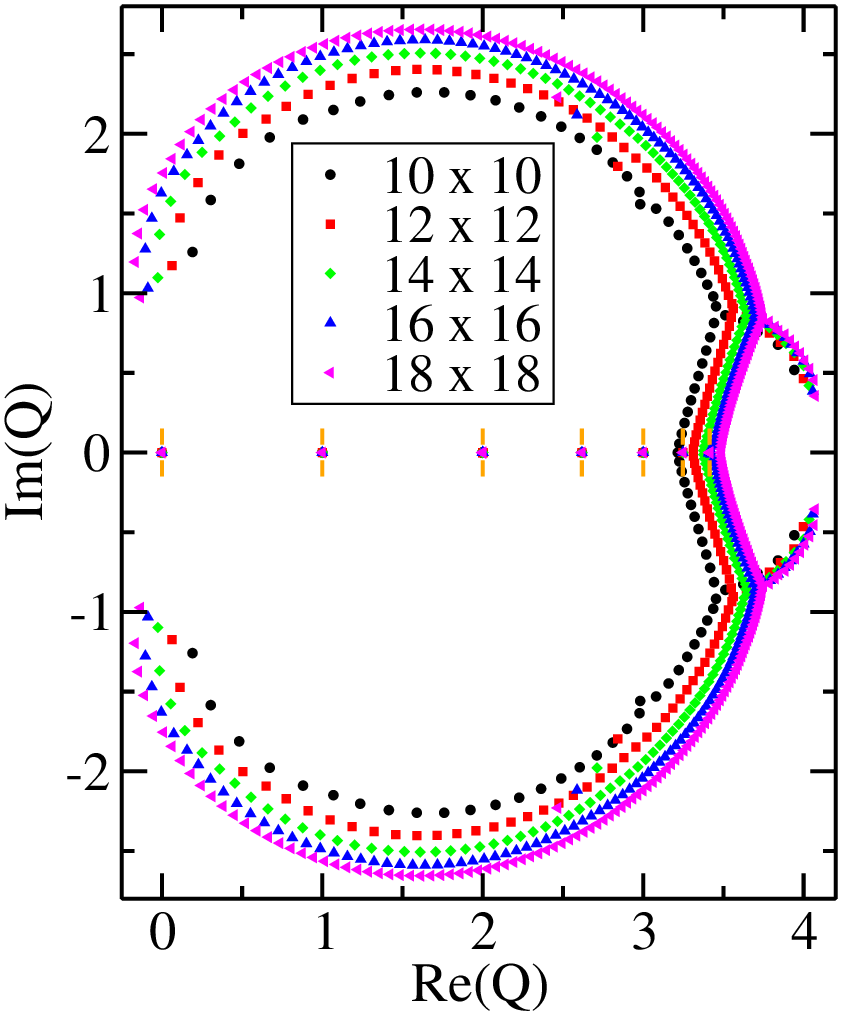}
\end{center}
\caption{Chromatic roots for $L \times L$ sections of the triangular
  lattice with free boundary conditions, and
  $L=10,12,14,16,18$. Beraha numbers $B_k$ with $k=2,3,4,5,6,7,8$ are
  shown as small vertical line segments.}
\label{fig:tri_latt}
\end{figure}

The resulting locations of the chromatic roots in the complex
$Q$-plane are shown in Fig.~\ref{fig:sq_latt} for the square lattice,
and in Fig.~\ref{fig:tri_latt} for the triangular lattice.  As in
earlier work \cite{paper1,paper2,paper3} we notice that most of the
roots fall on connected curves which, very roughly speaking, tend to
form a egg-shaped figure with a pair of prongs on the right side.
Without going into details (which are numerous, see
\cite{Baxter86,Baxter87,Saleur90,Saleur91,paper1,paper2,paper3,Baxter82}
for more exact statements) in the limit $L \to \infty$ the
egg-shaped part is expected to enclose a segment $Q \in [0,Q_0]$ of
the real axis, with $Q_0 = 3$ for the square lattice, and $Q_0
\approx 3.81967$ for the triangular lattice. In the interior of the
egg one observes additional isolated real zeros right at, or
extremely close to, the so-called Beraha numbers
\begin{equation}
 B_k = \big( 2 \cos (\pi/k) \big)^2 \qquad (k = 2,3,4,\ldots)
\end{equation}
The reason why this is so is linked to particularities of the
representation theory of the quantum algebra that underlies the
two-dimensional Potts model \cite{Saleur90,Saleur91}.

The existence of a finite-size equivalent of $Q_0$ is clearly visible
from Figs.~\ref{fig:sq_latt}--\ref{fig:tri_latt}. One could propose
defining $Q_0(L)$ as the largest real root, but one should be careful
since in some cases the egg-shaped curve does not possess a root
right on the real axis. Barring these (and other) difficulties, one
could speculate that if the result for an arbitrary planar graph was
qualitatively similar to that of these regular lattices, the
probability distribution of complex roots would look like some
broadened-out egg.  In addition, the distribution of real roots
would be a superposition of sharp peaks centered at the Beraha
numbers, and a broad background distribution corresponding to the
$Q_0(L)$.

\begin{figure}
\includegraphics[width=\columnwidth]{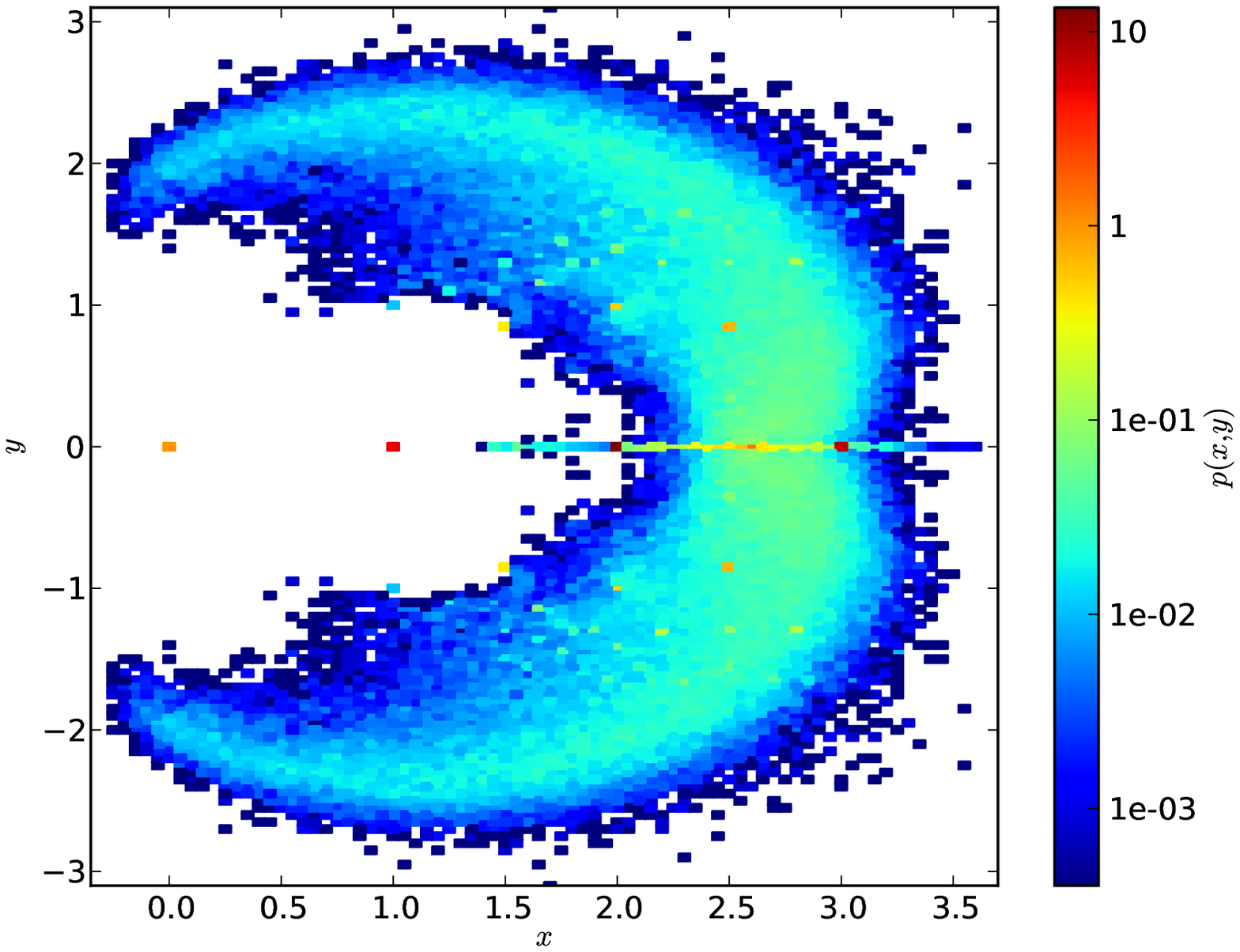}
\caption{Distribution for complex chromatic roots $Q = x + i y$ obtained from a
  sample of 2500 planar graphs of size 100. The normalisation is such
  that $p(x,y)$ is the expected number of roots in $[x - 0.05, x + 0.05]
  \times [y - 0.05, y + 0.05]$ counted with their multiplicity.}
\label{fig:roots_complex}
\end{figure}

\begin{figure}
\includegraphics[width=\columnwidth]{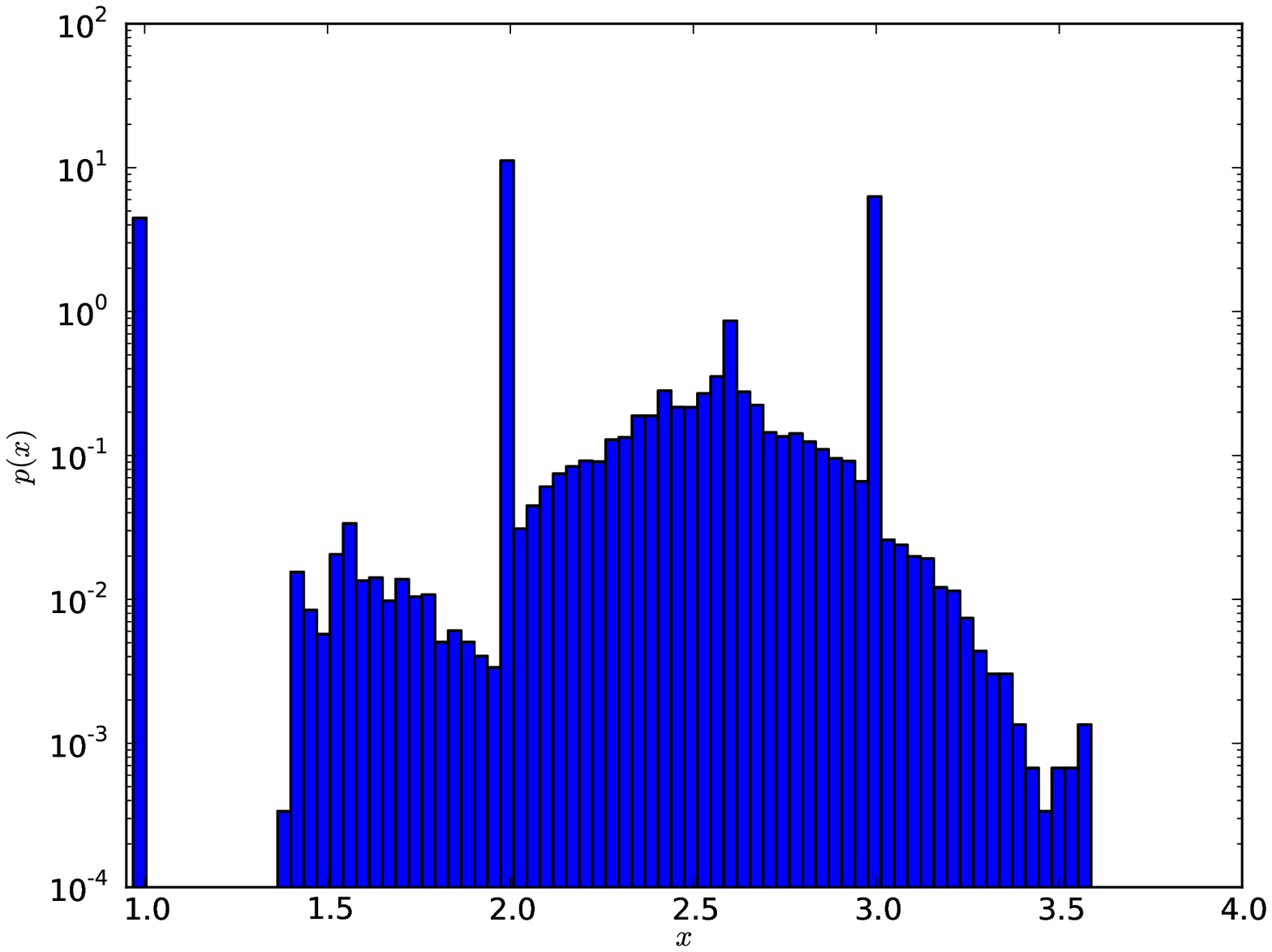}
\caption{Probability distribution for the subset of chromatic roots
  in Fig.~\ref{fig:roots_complex} that lie on the real
  axis. Normalised to unit total area.}
\label{fig:roots_real}
\end{figure}

This is indeed more-or-less what we observe. For a sample of 2500
uniformly drawn \cite{Fusy} planar graphs of $N=100$ vertices we show
the distributions of complex chromatic roots in
Fig.~\ref{fig:roots_complex}, and the distribution of the subset of
real roots in Fig.~\ref{fig:roots_real}. Regarding the complex roots,
we note that although chromatic roots of planar graphs have been shown
to be dense in the complex plane \cite{Sokal04} (except maybe in the
disc $|Q-1|<1$), typical roots are clearly quite close to the origin.

As to the real chromatic roots, the absense of roots on the negative
real axis, and the intervals $(0,1)$ and $(1,32/27]$ follow from a
theorem \cite{Jackson} (see also \cite{Thomassen}). The roots found
here respect this theorem as well as the Birkhoff-Lewis conjecture
\cite{BL46}.  Apart from that, Fig.~\ref{fig:roots_real} shows as
expected a superposition of a broad background distribution and sharp
peaks centered at $Q=B_k$ with $k=2,3,4,5,6$ (we have $B_5 \simeq
2.61803$).  It seems likely that for $N \to \infty$ this background
will extend to the interval $(32/27,4)$ and peaks will occur at all
Beraha numbers. We also expect that the maximum of the background
distribution may be shifted further to the right by requiring the
graphs to be two- or three-connected, and so presumably the peaks at
the first few $B_k$ should stand out more clearly.  We plan to
investigate this and further issues in more detail elsewhere
\cite{toappear}.

\section{Conclusion}
\label{sec:concl}

In this paper we have shown how to obtain an efficient algorithm for
solving instances of graph-related \#P-complete problems. The case of the
Potts partition function $Z_G(Q,v)$, and in particular its
specialisation the chromatic polynomial $\chi_G(Q) = Z_G(Q,-1)$, were
studied in detail, and some results on the distribution of real
and complex chromatic roots for uniformly drawn random planar graphs
were given.

Our algorithmic approach has some overlap with an algorithm
described---but apparently not implemented---by Noble \cite{Noble}.
This author also combines partial evaluations of $Z_G(Q,v)$---treated
by him in the Tutte polynomial formulation \cite{Tutte54}---in the
basis of partitions with the notion of tree decomposition. There are
however a number of important differences. First, our use of transfer
matrices---and specifically of its factorisation within each bag,
i.e., our choice to process a single edge at a time within each
bag---leads to a better time complexity. In particular, treating a
daughter bag with no sisters has worst-case running time $C_{n_{\rm
    max}}^2 n_{\rm max}^2 + C_{n_{\rm max}} 2^{n_{\rm max}(n_{\rm
    max}-1)/2}$ in Noble's approach versus $C_{n_{\rm max}} n_{\rm
  max}$ in ours. Second, we treat binary fusions more efficiently,
exploiting the fact that a given pair ${\cal P}_1$, ${\cal P}_2$ of
input partitions gives rise to a {\em unique} output partition ${\cal
  P}_1 \vee {\cal P}_2$: this reduces the time complexity for binary
fusions from worst-case time $C_{n_{\rm max}}^3 n_{\rm max}^2$ in
\cite{Noble} to $C_{n_{\rm max}}^2 n_{\rm max}$ in our
approach. Third, our algorithm appears easier to describe and
implement---as we have indeed done---and it avoids having to deal with
the $Q \to 0$ limit ($x=1$ in the notations of \cite{Noble}) as a
particular case.

\begin{figure}
\includegraphics[width=\columnwidth]{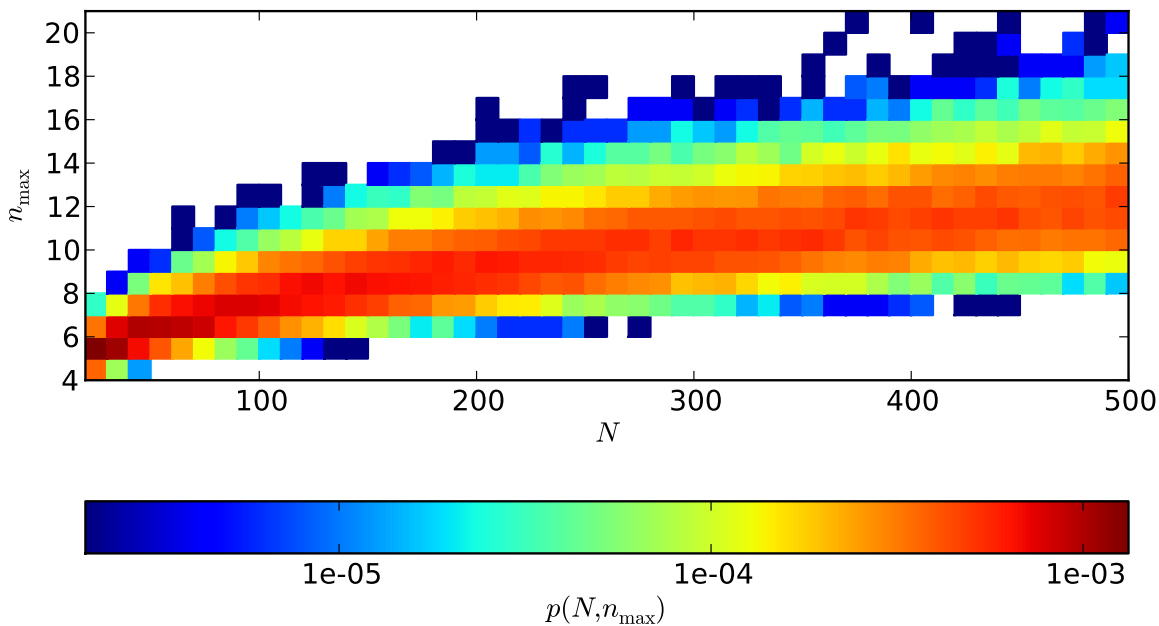}
\caption{Probability distribution $p(N, n_{\rm max})$ of obtaining a
  tree decomposition of maximum bag size $n_{\rm max}$ (i.e., width
  $n_{\rm max}-1$) when applying the {\tt GreedyFillIn} algorithm to
  random planar graphs, as a function of the graph size $N$. The data
  is computed over a sample of 100 planar graphs for each size between
  20 and 500, binned in blocks of 10 on the x axis and normalised to
  unit total area.}
\label{fig:tw}
\end{figure}

It is thus clear that what matters the most is to obtain a good tree
decomposition, and in particular the running time is essentially
determined by $n_{\rm max}$. In Fig.~\ref{fig:tw} we show the
distribution of $n_{\rm max}$ obtained from applying the algorithm
{\tt GreedyFillIn} to random planar graphs as a function of $N$.

It is plainly visible from Fig.~\ref{fig:tw} that both the mean
and worst-case $n_{\rm max}$ exhibit a slower than linear growth
with $N$. Assuming that {\tt GreedyFillIn} yields always an $n_{\rm max}$
which is of the order of the true tree width, we would have
$n_{\rm max} \le {\rm cst} \sqrt{N}$, and the data of Fig.~\ref{fig:tw}
appear to be compatible with such a scaling. We defer the further
investigation of the asymptotic behaviour of {\tt GreedyFillIn} to
future work \cite{toappear}.
  
We should also mention that the use of tree decomposition in the context
of {\em decision} problems has previously been advocated by a number of
authors (see \cite{Fomin1} and references therein). In these works, the
solution is found by doing dynamic programming on the tree-decomposed
graph.

Let us conclude by commenting on the generality of our algorithm.
Going through the list of known NP-complete problems related to graphs
and network design, one easily realises that most of them can be
promoted to counting problems (which cannot be easier), and a counting
polynomial analogous to $\chi_G(Q)$ can be defined. Adapting our
algorithm to those cases usually requires just a problem-specific
definition of the states, of the single-edge transfer process, and of
the fusion procedure, whereas the bulk of the method can be taken over
without changes.

It is worth underlining that our ability to solve efficiently an
enumeration problem can be exploited to obtain a specific instance
that solves the corresponding decision problem with only an additional
linear time factor.

In particular, we have adapted (but not yet implemented) our algorithm
to counting versions of the following problems: Hamiltonian walks and
cycles, longest path, vertex cover, dominating set, feedback vertex
set, and minimum maximal independent set. We plan to report on these
problems elsewhere \cite{toappear}.

\section*{Acknowledgments}

This work was supported by the European Community Network ENRAGE
(grant MRTN-CT-2004-005616) and by the Agence Nationale de la
Recherche (grant ANR-06-BLAN-0124-03). The authors thank A.D.\ Sokal
for discussions and for commenting on an earlier version of the
manuscript.

\end{document}